\documentclass[conference]{IEEEtran}
\IEEEoverridecommandlockouts
\usepackage{cite}
\usepackage{amsmath,amssymb,amsfonts}
\usepackage{algorithm}
\usepackage{algpseudocode}
\algnewcommand{\LineComment}[1]{\State \(\triangleright\) #1}
\usepackage{graphicx}
\usepackage{float}
\usepackage{subfigure}
\usepackage{textcomp}
\usepackage{xcolor}
\usepackage{threeparttable}
\def\BibTeX{{\rm B\kern-.05em{\sc i\kern-.025em b}\kern-.08em
    T\kern-.1667em\lower.7ex\hbox{E}\kern-.125emX}}
\begin{document}

\title{A Low-Power Sparse Deep Learning Accelerator with Optimized Data Reuse}

\author{\IEEEauthorblockN{Kai-Chieh Hsu and Tian-Sheuan Chang, \textit{Senior Member, IEEE}}
\IEEEauthorblockA{\textit{Institute of Electronics, National Yang Ming Chiao Tung University,} 
Hsinchu 300093, Taiwan \\
\IEEEauthorblockA{Email: hsukaij@gmail.com, tschang@nycu.edu.tw}
}
}
\maketitle

\begin{abstract}
Sparse deep learning has reduced computation significantly, but its irregular non-zero data distribution complicates the data flow and hinders data reuse, increasing on-chip SRAM access and thus power consumption of the chip. This paper addresses the aforementioned issues by maximizing data reuse to reduce SRAM access by two approaches. First, we propose Effective Index Matching (EIM), which efficiently searches and arranges non-zero operations from compressed data. Second, we propose Shared Index Data Reuse (SIDR) which coordinates the operations between Processing Elements (PEs), regularizing their SRAM data access, thereby enabling all data to be reused efficiently. Our approach reduces the access of the SRAM buffer by 86\% when compared to the previous design, SparTen. As a result, our design achieves a 2.5$\times$ improvement in power efficiency compared to state-of-the-art methods while maintaining a simpler dataflow.

\end{abstract}

\section{Introduction}

Deep learning's high computational demands lead to significant energy consumption and usage of hardware resources. To mitigate these issues, weight pruning\cite{han2015deep} and sparse computation have been introduced for deep learning accelerators (DLAs). However, this introduces irregular data access and computation, and thus two major design challenges: complex data flow and hardware overhead, and significant SRAM access due to low data reuse.

The first challenge is due to the matching of non-zero sparse input and weight indexes for non-zero multiply-accumulate operations. This irregularity makes this index matching consume significant power and area overhead~\cite{gondimalla2019sparten}. Besides, 
during the index matching process, the probability of successfully matching the corresponding non-zero operands decreases rapidly as sparsity increases. Thus, to prevent computation units from idle due to failed matching attempts, previous work~\cite{gondimalla2019sparten} uses multiple matching units, which incurs substantial hardware costs.

For the second challenge on SRAM access, a typical accelerator for dense computation can share its weight and input for all processing elements (PEs) and thus reduce SRAM data access. To analyze the required SRAM buffer bandwidth for a given dataflow, we define an indicator called the Memory Access per MAC (MAPM), which represents the average number of data bytes accessed from the on-chip SRAM per MAC operation. This indicator is expressed as bytes per MAC (byte/MAC). Take the multiplication of two dense 4$\times$4 matrices as an example. Assuming the inputs and outputs of the MAC units are 8 bits, if there is no data reuse, the MAPM would be as high as 4 byte/MAC (reading two operands for the multiplication, one for the addition, and writing back the result). In contrast, with full data reuse as in typical dense computing DLAs~\cite{chen2016eyeriss,chang2019vwa}, such as when using a 4$\times$4 output stationary systolic array, 32 input data are read from memory, 64 MAC operations are performed, and 16 output data are written back to memory, reducing the MAPM to 0.75 byte/MAC.

The irregular nature of the sparse data distribution makes it challenging to reuse all input and output data in PEs. For example, Sparten\cite{gondimalla2019sparten}, which adopts the dot product, reuses only the output data, while SCNN~\cite{parashar2017scnn}, which uses the Cartesian product, reuses only the input data. Their MAPM values are 2.09 byte/MAC and 2.03 byte/MAC, respectively. Since accessing SRAM consumes considerable energy, insufficient data reuse significantly increases SRAM buffer read/write operations, leading to a significant increase in total chip power consumption.

To address these challenges, we propose the \textbf{E}ffective \textbf{I}ndex \textbf{M}atching (\textbf{EIM}) and the \textbf{shared index data reuse} (\textbf{SIDR}). \textbf{EIM} re-sorts the bitmap representing the indexes of required data to the order of the compressed data, allowing it to determine the indexes of the required data within the buffer. \textbf{SIDR} merges non-zero index addresses from multiple PEs such that common index addresses will appear once to access the SRAM and reuse the accessed data for shared ones. With this approach, the hardware dataflow also becomes regular.
Thus, we propose a sparse accelerator based on dense structure. It uses dense DLA designs that broadcast input and weight for maximum data reuse, but accesses SRAM and activates PEs only for non-zero operations to save SRAM access and cycle count. This combines the benefits of data regularity of dense DLAs~\cite{chen2016eyeriss,chang2019vwa} and computing efficiency of sparse DLAs~\cite{gondimalla2019sparten,wang2021dual}. 

The rest of the paper is organized as follows. Section II presents our proposed approach. Section III shows the experimental results. Finally, this paper is concluded in Section IV.

\begin{figure}[t]
\centering
\includegraphics[width=0.4\textwidth]{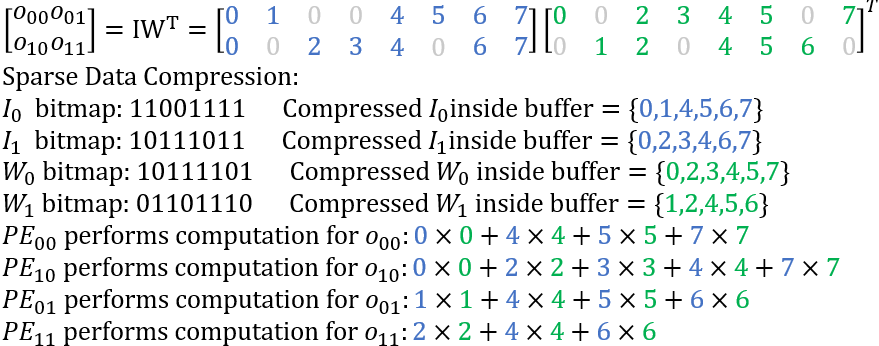}
\caption{An example of the data compression using bitmap. The blue and green numbers represent the original indexes of non-zero inputs and weights, respectively, while the gray zeros represent the zero values}
\label{Bitmap example}
\end{figure}

\begin{figure}[t]
\centering
\includegraphics[width=0.42\textwidth]{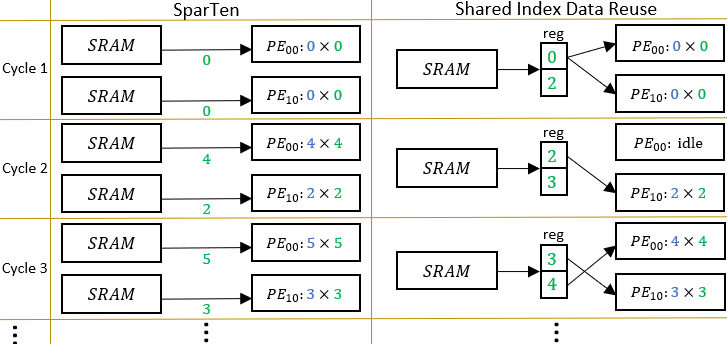}
\caption{A Simple Example for typical sparse DLAs~\cite{gondimalla2019sparten} and proposed SIDR}
\label{SIDR example}
\end{figure}

\section{Methodology}
\subsection{Overview of the approach}
Fig.~\ref{SIDR example} illustrates the computation of $o_{00}$ and $o_{10}$, as shown in Fig.~\ref{Bitmap example}, to demonstrate the typical flow in sparse DLAs~\cite{gondimalla2019sparten} and the proposed \textit{shared index data reuse}. We use the bitmap as the sparse data format to represent the original indexes of non-zero data, as illustrated in Fig.~\ref{Bitmap example}, while data with zero values are skipped and stored in the buffer.

In this example, a typical sparse DLA uses index matching to determine that $PE_{00}$ needs to read weights with original index 0, 4, 5, and 7 from buffers, while $PE_{10}$ needs to read weights with original index 0, 2, 3, 4, and 7 from buffers. The PEs sequentially read the weights from buffers according to these indexes and perform MAC operations. Since weights with original indexes 0, 4, and 7 are read from buffers twice, causing unnecessary SRAM access, these repeatedly accessed data can be shared and reused. To enable such reuse, we propose the SIDR. The concept of the SIDR employs the hierarchical memory to buffer a set of data (referred to as the shared data) by registers (referred to as the shared register). This can be managed by a shared index, and shared among multiple PEs. The PEs fetch data through multiplexers (MUX) controlled by offset index. As a result, all weights in SRAM are read only once. The details of the SIDR will be introduced in Section~\ref{Detail of SIDR}.

\subsection{Shared Index Data Reuse to 2-D Arrays}
\begin{figure}[t]
\centering
\includegraphics[width=0.47\textwidth]{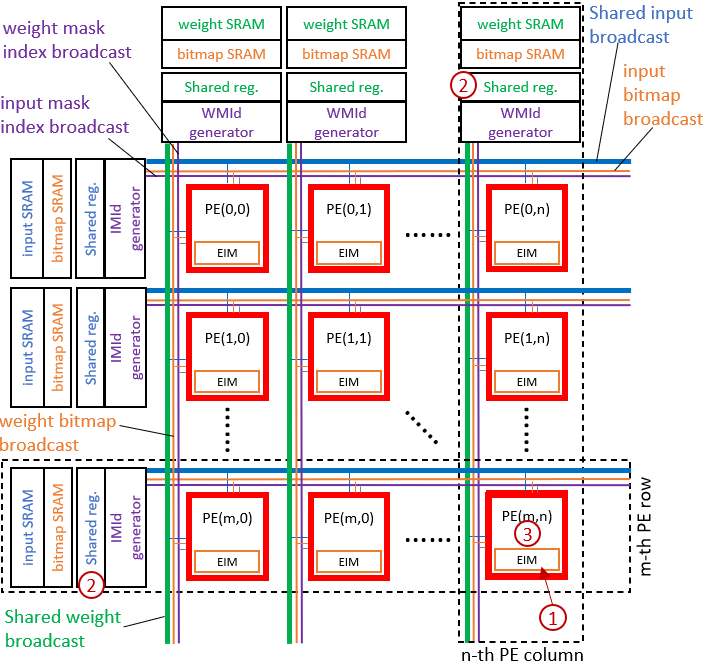}
\caption{The DLA architecture and the dataflow of SIDR.}
\label{DCMA process}
\end{figure}

Above \textit{shared index data reuse} can be applied to two PEs directly, but this is not efficient for massive computing of deep learning. A better way is to use 2-D PE arrays and apply \textit{SIDR} on both weight and input. A typical 2-D PE array like in~\cite{chang2019vwa}, as shown in Fig.~\ref{DCMA process}, broadcasts input and weights to all PEs for maximum data reuse. The PEs in the same row share the inputs by a shared register, while the PEs in the same column share the weights. Based on this flow, the \textit{shared index data reuse} can be divided into three steps. 
The first step is to determine which multiplication is a non-zero operation that PEs must execute and identify the index to access required input and weight from buffer, referred to as effective input index (\textit{EffI}) and effective weight index (\textit{EffW}) respectively. Fig.~\ref{SIMIM process} provides an example of the following effective index matching (EIM), which will be elaborated in Section~\ref{Detail of SIMIM}. 
The second step is to coordinate PEs and decide which data shall be buffered by the shared registers and broadcast. Section~\ref{Detail of SIDR} provide a detailed description of this step. 
The third step is to fetch data and perform the output-stationary MAC operation.

\subsection{Proposed Effective Index Matching (EIM) for Sparse Multiplications}
\label{Detail of SIMIM}
\begin{figure}[t]
\centering
\includegraphics[width=0.48\textwidth]{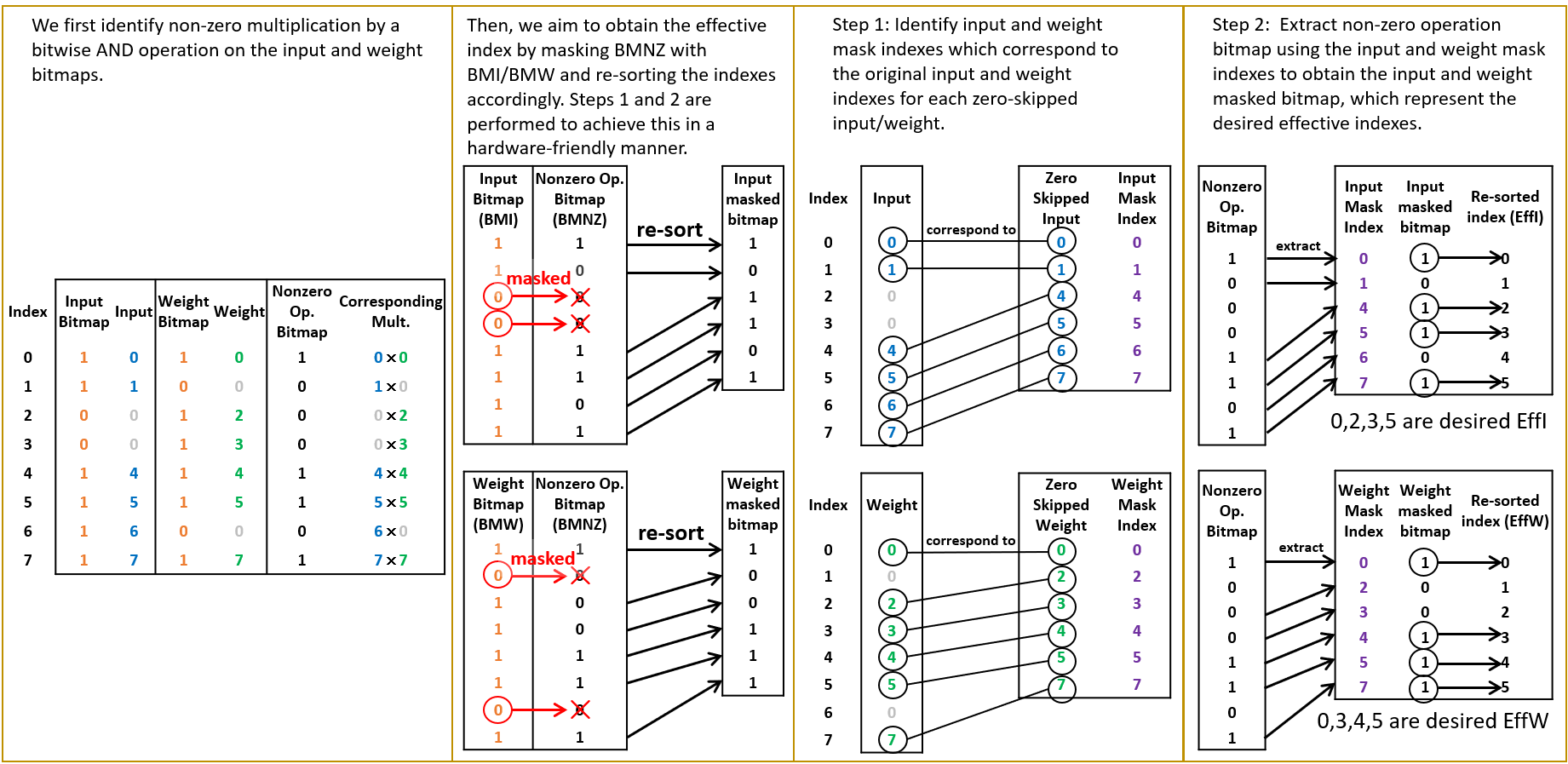}
\caption{The process and example of EIM.}
\label{SIMIM process}
\end{figure}

The effective index matching is to identify non-zero multiplications and their effective indexes at weight and input, \textit{EffI} and \textit{EffW}, in a regular way. These effective indexes enable the PEs to fetch data accordingly, allowing multiple matching attempts to occur simultaneously and preventing PEs from idle due to failed matches. Fig.~\ref{SIMIM process} illustrates how to get these effective indexes as shown in Fig.~\ref{Bitmap example}. 

An intuitive approach is to identify non-zero multiplications by a bitwise AND operation on the input and weight bitmaps, \textit{BMI} and \textit{BMW}, respectively to obtain the non-zero operation bitmap \textit{BMNZ}. 
Note that \textit{BMNZ} indicates non-zero multiplications but also represents the indexes of the data to be read. Thus, we can get the effective index by masking \textit{BMNZ} with \textit{BMI}/\textit{BMW} and re-sorting the indexes accordingly, resulting in the input masked bitmap (\textit{IMBM}) and the weight masked bitmap (\textit{WMBM}). However, this masking method is not hardware efficient due to irregular output order. 

Instead, we use the following two steps to generate these effective indexes.
First, we identify the original index of input and weight corresponding to each re-sorted index, referred to as input mask index and weight mask index, \textit{IMId} and \textit{WMId} respectively. Since each row and column of PEs uses identical input and weight bitmap with the same re-sorting order, these PEs share the input and weight mask indexes, respectively, thereby reducing the hardware overhead. 
Second, we extract non-zero operation bitmap using the input and weight mask index to obtain the input masked bitmap and weight masked bitmap, which represent the desired effective indexes. Finally, these effective indexes are sequentially pushed into the FIFOs, $EIM\_FIFO_I$ and $EIM\_FIFO_W$, which serve as buffer between the index matching unit and MAC.

\subsection{Detailed Dataflow of Shared Index Data Reuse}
\label{Detail of SIDR}
\begin{figure}[t]
\centering
\includegraphics[width=0.48\textwidth]{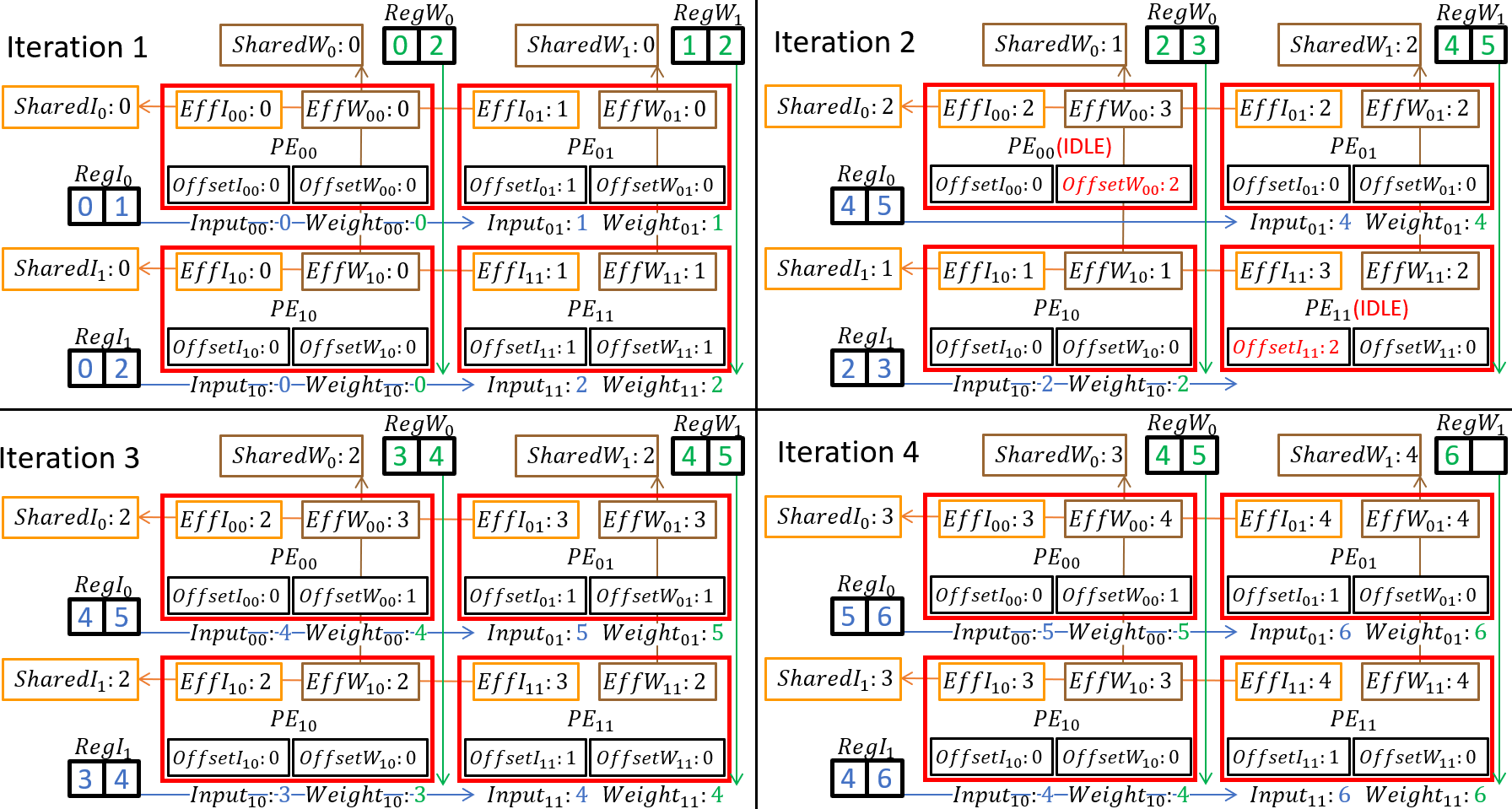}
\caption{The demonstration of performing the example in Fig.~\ref{Bitmap example} using SIDR}
\label{2D SIDR example}
\end{figure}

The SIDR procedure is illustrated in Algorithm~\ref{SIDR pseudocode}, with an example provided in Fig.~\ref{2D SIDR example}.
SIDR buffers a range of data, enabling data sharing among multiple PEs. This method requires coordinating the PEs to ensure that their execution progress remains nearly synchronized, so the data they need is located at nearby indexes within the buffer. To achieve this, PEs that are lagging behind are given higher priority for execution, while PEs that are too far ahead are made to wait. The detailed operation is divided into the following five steps.

First, each PE obtains the effective input and weight index for upcoming multiplication by performing EIM.
Second, the system determines the shared input and weight indexes, $SharedI_m$ and $SharedW_n$, for each row and column of PEs to decide which input and weight should be buffered by the shared resister and broadcasted. To prevent lagging PEs from idle, the shared input and weight indexes are set to the smallest effective input and weight indexes within the same row and column, ensuring that these PEs can successfully fetch the required data.
Third, the shared input and weight registers, $RegI_m$ and $RegW_n$, buffer data from the input and weight buffers, $BufI_m$ and $BufW_n$, based on the shared input and weight indexes, and then broadcast the shared input and weight.
Fourth, each PE identifies the indexes of the required input and weight in the shared register, $OffsetI$ and $OffsetW$ respectively, which are the differences between the effective indexes and the shared indexes.
Finally, if both the required input and the weight for a PE are available in the shared register, the PE then fetches them to carry out the output-stationary MAC operation.
The system repeats these steps until all operations are fully completed.

\begin{algorithm}
\caption{SIDR for a 16 $\times$ 16 2-D PE array with Shared Registers size of 8}\label{SIDR pseudocode}
\begin{algorithmic}

\While{MACs are not completed} \Comment{Iterate sequentially}

\LineComment{Obtain Effective Index for upcoming operation}
\For {each $m,n$ in PE Array} \Comment{Hardware parallelism}
\If{$PE(m,n)$ is not IDLE in last iteration}
    \State $EffI_{m,n} \gets pop(EIM\_FIFO_I(m,n))$
    \State $EffW_{m,n} \gets pop(EIM\_FIFO_W(m,n))$
\EndIf
\EndFor

\LineComment{Determine $SharedI$ to buffer Shared Input}
\For {$m < 16$} \Comment{Hardware parallelism}
    \State $SharedI_m \gets min(EffI_{m,j}$, for each $j < 16)$ %
    \State $RegI_m \gets BufI_m[SharedI_m:SharedI_m+7]$
\EndFor

\LineComment{Determine $SharedW$ to buffer Shared Weight}
\For {$n < 16$} \Comment{Hardware parallelism}
    \State $SharedW_n \gets min(EffW_{i,n}$, for each $i < 16)$ %
    \State $RegW_n \gets BufW_n[SharedW_n:SharedW_n+7]$
\EndFor

\LineComment{Fetch Shared Data and execute MAC operation}
\For {each $m,n$ in PE Array} \Comment{Hardware parallelism}
\State $OffsetI_{m,n} \gets EffI_{m,n} - SharedI_m$
\State $OffsetW_{m,n} \gets EffW_{m,n} - SharedW_n$
\If{$OffsetI_{m,n} < 8 \And OffsetW_{m,n} < 8$}
    \State $I_{m,n} \gets RegI_m[OffsetI_{m,n}]$
    \State $W_{m,n} \gets RegW_n[OffsetW_{m,n}]$
    \State $MAC_{m,n}$ Accumulate $ I_{m,n} \times W_{m,n}$
\Else
    \State $PE_{m,n}$ IDLE
\EndIf
\EndFor

\EndWhile
\end{algorithmic}
\end{algorithm}

\begin{table}[]
\caption{Comparison with previous works}
\label{TABLE}
\begin{threeparttable}
\begin{tabular}{|l|l|l|l|l|l|l|}
\hline
\multicolumn{1}{|c|}{}                                                  & \multicolumn{1}{c|}{\begin{tabular}[c]{@{}c@{}}SparTen\\ ~\cite{gondimalla2019sparten}\end{tabular}} & \multicolumn{1}{c|}{\begin{tabular}[c]{@{}c@{}}Eyeriss \\ v2\\ ~\cite{chen2019eyeriss}\end{tabular}} & \multicolumn{1}{c|}{\begin{tabular}[c]{@{}c@{}}SIGMA\\ ~\cite{qin2020sigma}\end{tabular}} & \multicolumn{1}{c|}{\begin{tabular}[c]{@{}c@{}}SNAP\\ 65nm-8b\\ ~\cite{zhang2020snap}\end{tabular}} & \multicolumn{1}{c|}{\begin{tabular}[c]{@{}c@{}}ORSAS\\ ~\cite{lin2023orsas}\end{tabular}} & \multicolumn{1}{c|}{\begin{tabular}[c]{@{}c@{}}Our \\ work\end{tabular}} \\ \hline
Technology                                                              & 45nm                                                                                                 & 65nm                                                                                                 & 28nm                                                                                      & 65nm                                                                                                & 55nm                                                                                      & 28nm                                                                     \\ \hline
Precision                                                               & -                                                                                                     & fxp8                                                                                                 & bfp16                                                                                     & fxp8                                                                                                & fxp8                                                                                      & fxp8                                                                     \\ \hline
\# of MACs                                                              & 32                                                                                                   & 384                                                                                                  & 16384                                                                                     & 252                                                                                                 & 256                                                                                       & 256                                                                      \\ \hline
\begin{tabular}[c]{@{}l@{}}Clock\\ Frequency\\ (Hz)\end{tabular}        & 800M                                                                                                 & 200M                                                                                                 & 500M                                                                                      & 250M                                                                                                & 200M                                                                                      & 800M                                                                     \\ \hline
\begin{tabular}[c]{@{}l@{}}Throughput\\ (TOPS)\end{tabular}             & 0.05                                                                                                 & 0.07$^\ddagger$                                                                                                     & 10.8                                                                                      & 0.126                                                                                               & 0.102                                                                                     & 0.27                                                                     \\ \hline
Area ($mm^2$)                                                           & 0.766                                                                                                & -                                                                                                     & 65.1                                                                                      & 9.32                                                                                                & 7.5                                                                                       & 0.926                                                                    \\ \hline
\# of  Gates                                                            & -                                                                                                     & 2695k                                                                                                & -                                                                                          & -                                                                                                    & -                                                                                          & 438k                                                                     \\ \hline
Power (W)                                                               & 0.118                                                                                                & 0.57                                                                                                 & 22.33                                                                                     & 0.5                                                                                                 & 0.198                                                                                     & 0.231                                                                    \\ \hline
\begin{tabular}[c]{@{}l@{}}Energy \\ Efficiency\\ (TOPS/W)\end{tabular} & 0.43$^\dagger$                                                                                                 & 0.251$^\ddagger$                                                                                                & 0.48                                                                                      & 0.25$^\dagger$                                                                                                & 0.52$^\dagger$                                                                                      & \begin{tabular}[c]{@{}l@{}}1.198\\ 2.066$^\dagger$\end{tabular}                    \\ \hline
\end{tabular}
    \begin{tablenotes}\footnotesize
    \item $\dagger$  Energy efficiency are measured under the assumption of 100\% PE utilization.
	\item $\ddagger$ TOPS include zero multiplication.
    \end{tablenotes}
\end{threeparttable}
\end{table}

\begin{figure}[t]
\begin{center}
\includegraphics[width=0.42\textwidth]{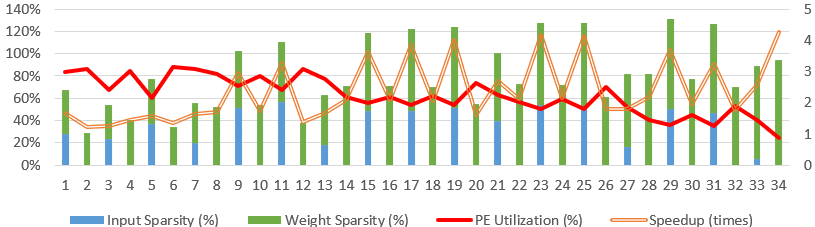}
\caption{The PE utilization and speedup for each PW layer of Mobilenetv2}
\label{mobilenet result}
\end{center}
\end{figure}

\section{Experimental Result}

To evaluate our proposed method, we developed a deep learning accelerator with a 16$\times$16 2-D PE array to execute sparse matrix operations using EIM and SIDR, as shown in Fig.~\ref{DCMA process}. The design is synthesized with TSMC's 28nm technology, runs at 0.9V, and has a clock rate at the 800MHz. Each PE includes an 8-bit fixed-point multiplier and a 24-bit fixed-point adder, with each shared register capable of buffering 8 inputs or weights. The total area of our design is $0.926 mm^2$, equivalent to 438K gate count.

\subsection{Experimental Result for PW Layer of MobileNet V2}
The following shows the performance evaluation results of the proposed method. We conducted inference on ImageNet~\cite{deng2009imagenet} using unstructured sparse MobileNet V2~\cite{sandler2018mobilenetv2}, where 75\% of its weights were pruned through the global L1 fine-grained pruning~\cite{han2015deep}. The results of each pointwise convolution (PW) layer and comparison with previous work are shown in Fig.~\ref{mobilenet result} and Table \ref{TABLE}. Our overall PE utilization reached 66\%, achieving a 2.1$\times$ speedup. The average MAPM of our design is just 0.29 byte/MAC, representing an 86\% reduction compared to SparTen~\cite{gondimalla2019sparten}. By reducing memory access, we achieved an overall energy efficiency of 1.2 TOPS/W, representing a 2.5$\times$ improvement compared to SIGMA~\cite{qin2020sigma}. It is noteworthy that energy efficiency measurements vary across studies. For example, Eyeriss v2~\cite{chen2019eyeriss} reports TOPS that include skipped zero computations, while SNAP~\cite{zhang2020snap} and ORSAS~\cite{lin2023orsas} measure energy efficiency under the assumption of 100\% PE utilization. We adopted the most rigorous method, as used by SIGMA, where TOPS only reflects actual non-zero computations performed by the hardware, with measurements taken under realistic conditions where PE utilization is not fully maximized in sparse computations. Additionally, under 100\% PE utilization, as seen in dense computations, our design can reach up to 2.1 TOPS/W.

\subsection{Experimental Result for Random Matrix Multiplication}
\begin{figure}[t]
\centering
\subfigure[Speedup]{
\label{random speedup}
\includegraphics[width=0.193\textwidth]{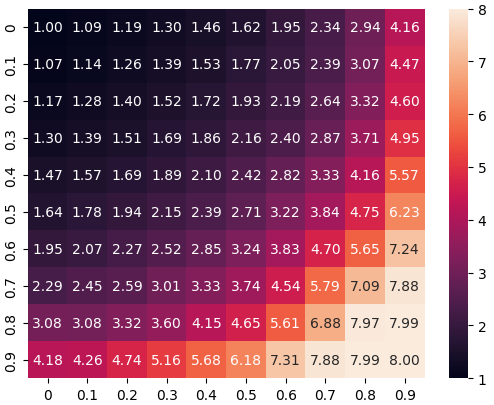}}
\subfigure[PE Utilization in Percentage]{
\label{random utilization}
\includegraphics[width=0.2\textwidth]{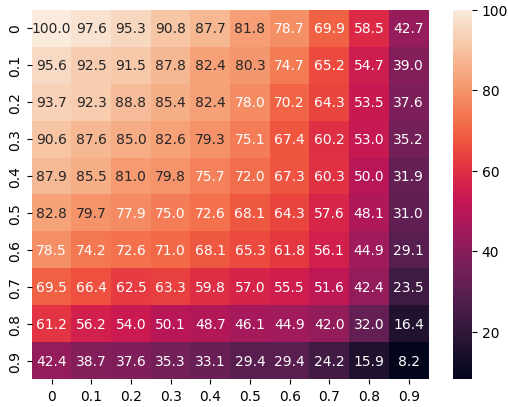}}
\caption{Experimental result for random matrix multiplication. The horizontal axis and vertical axis represent the sparsity of the input and weight matrices, respectively.}
\label{random result}
\end{figure}

We generated random 1024$\times$1024 matrices, then pruned them to various sparsity levels and carry out matrix multiplications to analyze the performance across different input and weight sparsity combinations. The results are shown in Fig.~\ref{random result}. In typical model inference, input and weight sparsity generally range between 50\% and 70\%. Within this range, our design maintains an average utilization rate of over 50\%, along with substantial acceleration.

\subsection{Power and Area Breakdown}

\begin{figure}[t]
    \begin{minipage}[t]{0.5\linewidth}
        \centering
        \includegraphics[width=0.8\textwidth]{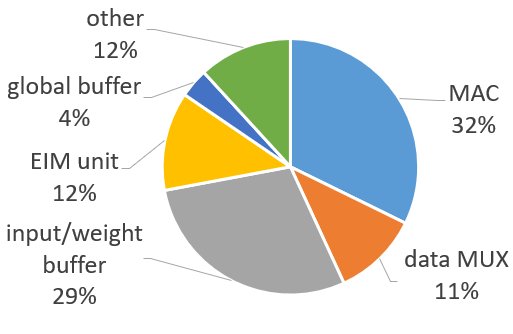}
        \caption{Power Breakdown}
        \label{power breakdown}
    \end{minipage}%
    \begin{minipage}[t]{0.5\linewidth}
        \centering
        \includegraphics[width=0.8\textwidth]{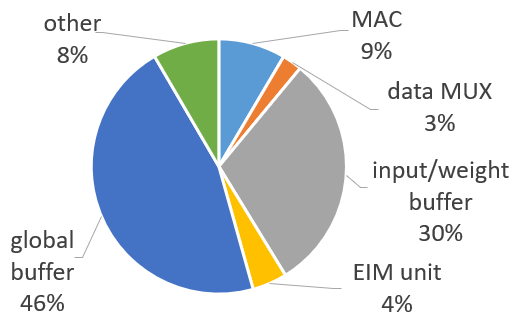}
        \caption{Area Breakdown}
        \label{area breakdown}
    \end{minipage}
\end{figure}

Figs.~\ref{power breakdown} and~\ref{area breakdown} illustrate the breakdown of power and area. It is evident that EIM incurs power and area overheads that are less than half of those of the MAC. Thanks to SIDR, the buffers usually remain in standby mode, which makes its power consumption proportionally much lower relative to its area.

\section{Conclusion}

In this work, we proposed an efficient sparse deep learning accelerator with complete data reuse, optimizing processing efficiency for sparse computations through effective index matching (EIM) and shared index data reuse (SIDR). EIM effectively identifies non-zero operations within compressed data with an acceptable overhead. SIDR, on the other hand, coordinates operations across PEs and maximizes data reuse. Our approach demonstrates a remarkable reduction in SRAM buffer access which not only minimizes total power consumption but also significantly improves power and area efficiency. These results validate the effectiveness of our approach, underscoring its potential for high-efficiency, low-power applications in sparse computation.

\bibliographystyle{IEEEtran}
\bibliography{citation.bib}

\end{document}